# Pressure screening in the interior of primary shells in double-wall carbon nanotubes

J. Arvanitidis[1,a)], D. Christofilos[1], K. Papagelis[2], K. S. Andrikopoulos[1], T. Takenobu[3], Y. Iwasa[3], H. Kataura[4], S. Ves[5], and G. A. Kourouklis[1]

[1]Physics Division, School of Technology, Aristotle University of Thessaloniki, 541 24 Thessaloniki, Greece

[2]Material Science Department, University of Patras, 265 04 Patras, Greece

[3]Institute for Materials Research, Tohoku University, 2-1-1 Katahira, Aoba-ku, Sendai 980-8577 and CREST, Japan Science and Technology Corporation, Kawaguchi 332-0012, Japan

[4]National Institute of Advanced Industrial Science and Technology (AIST), 1-1-1 Higashi, Tsukuba, Ibaraki 305-8562, Japan

[5]Physics Department, Aristotle University of Thessaloniki, 541 24 Thessaloniki, Greece

The pressure response of double-wall carbon nanotubes has been investigated by means of Raman spectroscopy up to 10 GPa. The intensity of the radial breathing modes of the outer tubes decreases rapidly but remain observable up to 9 GPa, exhibiting a behavior similar (but less pronounced) to that of single-wall carbon nanotubes, which undergo a shape distortion at higher pressures. In addition, the tangential band of the external tubes broadens and decreases in amplitude. The corresponding Raman features of the internal tubes appear to be considerably less sensitive to pressure. All findings lead to the conclusion that the outer tubes act as a protection shield for the inner tubes whereas the latter increase the structural stability of the outer tubes upon pressure application.

Carbon nanotubes have attracted intense scientific interest due to their fascinating essentially one-dimensional electronic and vibrational band structure, their unique mechanical properties as well as the prospect for numerous applications. Raman spectroscopy has become a widespread tool for the analysis and characterization of carbon nanotubes and numerous high-pressure Raman scattering studies, on single-wall carbon nanotubes (SWCNTs) and multi-wall carbon nanotubes, have made important contributions towards the understanding of the physical properties of these materials.[1-3] A discontinuous reduction in the intensity of the low frequency radial breathing modes (RBMs) near 2 GPa has been observed in SWCNTs, accompanied, in some cases, by changes in the pressure coefficients of the tangential modes to lower values. These experimental findings have been attributed to a pressure induced hexagonal[1] or oval[2] distortion of the cylindrically shaped cross section of the bundled nanotubes. In addition, high-pressure x-ray diffraction (XRD) measurements together with theoretical calculations suggest a structural distortion at ~1.5 GPa, which is also associated with a pressure-induced nanotube polygonization.[4] Raman spectroscopy at ambient pressure has been also successfully employed in the study of the more recently observed[5] and synthesized in bulk quantities[6] double-wall carbon nanotubes (DWCNTs), suggesting that the outer tubes provide an unperturbed environment to their interior[7] and that the interaction in a DWCNT bundle is stronger than the inner-outer tube interaction.[8] In this work, we study the effect of high-pressure on DWCNTs by means of Raman spectroscopy in order to investigate their structural stability and compare it with that of SWCNTs, elucidating the differences induced by the inner-outer tube interaction.

The starting raw SWCNT material was generated by the pulsed laser vaporization of a carbon rod with Ni and Co catalysts in a furnace operated at 1473 or 1523 K.[9] The peapods, prepared by a reaction of the purified uncapped SWCNTs with $C_{60}$ vapor,[10,11] were converted into bundled DWCNTs by heating for 5 h at 1473 K in vacuum, following Bandow's procedure.[6] Raman spectra of the DWCNTs were recorded in the back-scattering geometry using a micro-Raman, triple grating system (DILOR XY) equipped with a cryogenic CCD detector. The spectral resolution of the system was ~2 cm$^{-1}$. High pressure Raman measurements were carried out using a Mao-Bell type diamond anvil cell (DAC). The 4:1 methanol-ethanol mixture was used as pressure transmitting medium and the ruby fluorescence technique was used for pressure calibration. For excitation, the 514.5 nm line of an Ar$^+$ laser was focused on the sample by means of a 20x objective, while the laser power was kept below 2.5 mW - measured directly before the cell - in order to eliminate laser-heating effects on the probed material and the concomitant softening of the observed Raman peaks.[12,13] The phonon frequencies were obtained by fitting Lorentzian functions to the experimental peaks, whereas numerical integration after background subtraction was used for the calculation of integrated intensities of the RBM bands.

Raman spectra of the DWCNT material at room temperature and various pressures up to 10.3 GPa are illustrated in Fig. 1. A spectrum taken after pressure release to ambient conditions is also included (top panel). Two frequency regions are displayed: i) 100-500 cm$^{-1}$, containing the RBMs of the carbon nanotubes and ii) 1300-1800 cm$^{-1}$, where the tangential modes of the rolled graphene sheets are located. It is well-known that in SWCNTs the frequencies of the RBMs, $\omega_{RBM}$ are inversely proportional to the diameter, $d_t$ of the tubes,[14] following the general expression $\omega_{RBM}$ (cm$^{-1}$)= $A/d_t$(nm) + $B$. For a rough estimation of the tube diameters, we have used the values $A$= 234 cm$^{-1}$·nm and $B$= 10 cm$^{-1}$, applied previously for SWCNT [15] and DWCNT [7] bundles.

Three main radial bands are observed at ambient conditions, each comprising of several individual Raman peaks, reflecting tubes with different chiral vector and the inner-outer tube interaction.[7,8] Two peaks located at 175 and 186 cm$^{-1}$ (labeled, as R$_1$ and R$_2$ in Table I) constitute the first RBM band. By means of the above-mentioned expression, these peaks are associated with carbon tubes of relatively large diameters, in the range 1.33-1.42 nm (outer or primary tubes). The second radial band extends from 300 to 350 cm$^{-1}$ comprising of at least three peaks, with the strongest located at ~323 cm$^{-1}$ ($R_4$), whereas the higher energy RBM band spans the frequency range 370-400 cm$^{-1}$. In the latter RBM band, four Raman peaks can be clearly resolved, with the one at ~384 cm$^{-1}$ ($R_5$) being intense and extremely narrow. Both bands containing the strong peaks $R_4$ and $R_5$ are attributed to the inner (secondary) nanotubes. Their frequencies suggest diameters in the range of 0.66-0.77 nm and 0.61-0.64 nm, respectively. The difference in the mean diameter of the primary and secondary tubes is ~0.7 nm, marginally larger than the double of the turbostratic constraint of graphite at room temperature (0.344 nm). These results are compatible with the XRD studies of a DWCNT material prepared with exactly the same method, revealing that the mean primary tube diameter is ~1.38 nm with an inner-outer tube separation of 0.36 nm.[16] In addition to the main three RBM bands, two weak and broad peaks are also resolved in the low frequency region at ~106 and ~267 cm$^{-1}$. The former could be assigned to carbon nanotubes with a very large diameter (~2.44 nm), while the latter corresponds to tubes of ~0.91 nm in diameter.

In the high frequency region of the Raman spectrum for the DWCNTs two main bands are observed at ambient conditions. The weaker band marked by "D" is attributed to a disorder-induced mode,[3,17] which also appears in graphite.[18] In our case, it is comprised mainly of two peaks located at 1320 ($D_1$) and 1348 cm$^{-1}$ ($D_2$). As the D band frequency also exhibits a downshift with decreasing nanotube diameter,[19] the $D_1$ and $D_2$ peaks should be ascribed to secondary tubes and primary tubes, respectively. The stronger Raman band marked by "G" is related to the $E_{2g}$ mode of graphite[18] and corresponds to in-plane carbon stretching vibrations in nanotubes (tangential band).[3,14] In SWCNTs, the tangential band contains two main components resulting from the carbon displacements parallel and perpendicular to the tube axis, usually labeled as $G^+$ and $G^-$, respectively.[3,15] Moreover, according to theoretical calculations,[17] the tangential band is expected to show a red shift for sufficiently small nanotube diameters, providing an additional splitting of this band in SWCNTs and DWCNTs.[20] In the DWCNT material investigated here, six components ($G_1$-$G_6$) are resolved at ambient pressure and their frequencies are tabulated in Table I. Based on previous Raman studies of SWCNTs and DWCNTs,[3,20] we interpret the $G_6$ strong Raman peak and the $G_5$ shoulder-like peak as the $G^+$ component of the carbon nanotubes, reflecting the existence of the primary and secondary tubes, respectively. The remaining $G_1$-$G_4$ peaks are attributed to the $G^-$ component of the various tubes having different size, keeping in mind that the lower energy $G^-$ peaks are associated with carbon nanotubes of smaller diameter.[21]

Upon pressure application all the observed Raman peaks shift towards higher energies, while at the same time significant relative intensity changes take place. The pressure dependence of the most characteristic Raman lines is illustrated in figure 2, while their pressure coefficients (parabolic when applicable) are given in Table I. With increasing pressure, the RBM band of the outer tubes ($R_1$ and $R_2$) displays strong intensity attenuation. Above 3 GPa the $R_1$ shoulder-like peak cannot be resolved from $R_2$, which disappears completely for pressures higher than 9 GPa. On the other hand, the RBM bands of the secondary tubes are hardly affected by the pressure, especially the $R_5$ peak of the small inner tubes, which remains narrow up to 10.3 GPa. This effect is quantitatively illustrated in figure 3(a), where the integrated intensities of the outer ($R_1$-$R_2$) and the larger inner tubes ($R_4$) RBM bands normalized to that of the smaller ones ($R_5$) are plotted against pressure. The relative integrated intensity of the $R_4$ band remains almost unaffected up to the highest pressure attained, in contrast to that of the outer tube RBM band, which decreases rapidly by an order of magnitude up to ~2.5 GPa. However, as already mentioned above, the $R_2$ peak, although weak, persists for pressures up to 9 GPa in contrast to the situation encountered in SWCNTs studies, where the RBM bands disappear above 1.5 or 1.7 GPa.[1,2]

It becomes evident from figure 2, that the $R_2$ peak exhibits a small sublinear behavior, similar to that predicted theoretically by Venkateswaran et al. for the RBM band in SWCNTs under high pressure.[1] According to their model, there is no penetration of the pressure transmitting medium into the interstitial channels of the nanotube bundles and the applied pressure causes a hexagonal distortion of the tube cross-section, eliminating the radial band. This description can be also adopted for the outer tubes in DWCNTs, although in this case the distortion of the outer tubes is expected to be smaller (the $R_2$ band persists with pressure), possibly due to the presence of the inner tubes. This assumption is further supported by the smaller pressure slopes of the $R_1$ and the $R_2$ peaks (primary tubes) in comparison to those reported in the literature for the RBM band of SWCNTs and compiled in ref. 3. Moreover, the pressure coefficient for the Raman peaks associated with the outer tubes is much larger than those corresponding to the inner ones (Fig. 2, Table I). The overall behavior of the RBM bands under pressure indicates that the outer carbon nanotubes are, by far, more vulnerable to pressure application than the inner tubes in line with the proposed stronger inter-DWCNT interaction than that between inner and outer tube at ambient conditions.[8] It seems logical to suggest that the existence of the primary tubes results in a screening of the applied pressure on the secondary tubes, while the latter provide structural support against pressure induced deformation of the outer tubes. Finally, the puzzling pressure response of the $R_3$ radial peak, corresponding to tubes of an intermediate diameter, must be noted. Namely, the peak intensity decreases with increasing pressure and its pressure evolution cannot be followed beyond 3 GPa, in close analogy to what is observed for the RBM band of the primary tubes. However, the very small pressure coefficient of the $R_3$ peak –comparable to those of the secondary tubes – prevents an unambiguous assignment.

The pressure response of the tangential band is also of great importance, further supporting the above considerations. The $G_6$ peak assigned to the $G^+$ band of the outer tubes exhibits a much larger pressure coefficient than that of $G_5$ attributed to the inner tubes, in agreement with our proposed assignment and the pressure screening effect inside the primary tubes. The different pressure coefficients result in a more clear separation of the two peaks at elevated pressures (Fig. 1). At the same time, a significant broadening and amplitude drop of the $G_6$ peak take place with pressure. In figure 3(b), the full width at half maximum (FWHM) of the $G_5$ and $G_6$ peaks is plotted as a function of pressure. It is evident that $G_6$ broadens much faster than $G_5$ peak, reflecting again the larger deformation of the outer tubes and the pressure screening for the inner ones. Another noticeable point is the sublinear pressure dependence of the $G_6$ peak position (similar to that of the corresponding radial band, $R_2$) in contrast to the superlinear behavior of $G_5$. This can be understood by assuming that with increasing pressure the inner-outer tube interaction becomes progressively stronger supporting the primary tubes (reduced slope), while at the same time the secondary tubes are increasingly affected by pressure (increased slope). The $G_4$ shoulder, assigned to the $G^-$ component of the larger primary tubes, shifts swiftly with pressure (like the $G^+$ component of these tubes) and merges with the $G_5$ and $G_6$ peaks above ~1 GPa. The rest of the G-peaks, associated with smaller nanotubes, display pressure coefficients considerably smaller than that of the $G_6$ peak and comparable with that of $G_5$. Their superlinear trend with pressure further supports their assignment to the secondary tubes.

The pressure dependence of the D band could not be followed at low pressures due to the overlap with the strong Raman signal of the diamond in the DAC around 1332 cm$^{-1}$. Only above 6 GPa, a weak and broad peak ($D_2$) appears in the measured spectral window. Its pressure behavior appears again to be sublinear in agreement with our tentative assignment of this peak to the outer shells. As peaks associated with inner tubes have much smaller slopes, the $D_1$ peak does not appear in our spectral window up to 10.3 GPa. Note that the broad and weak band observed at ~1455 cm$^{-1}$ inside the DAC (asterisk in Fig. 1), is absent in spectra taken outside the cell. This peak is attributed to the hydrostatic pressure medium of methanol-ethanol.[22]

Although the pressure-induced shifts of the Raman peaks in DWCNTs are fully reversible, this is not the case for the relative intensities of certain bands. Namely, the integrated intensity of the $R_1$-$R_2$ band and the amplitude of the $G_6$ peak do not fully recover after total pressure release. Moreover, the intensity of the D bands after pressure release remains somewhat larger to that initially recorded at ambient conditions. These divergences suggest the existence of residual pressure-induced deformations of the primary tubes, in analogy to those observed in SWCNTs.[1]

Summarizing, our high pressure Raman study on the DWCNTs show that the application of pressure initially causes the deformation of the primary tubes, which actually shield the inner tubes against pressure. At higher pressure, the increased interaction between outer and inner shells acts as to provide structural support against the deformation of the outer tubes.

This work was partly supported by NEDO and MEXT, Japan.

# References

[a] Author to whom correspondence should be addressed; electronic mail: jarvan@vergina.eng.auth.gr

**Figure Captions**

**Figure 1.** Raman spectra of the DWCNTs at room temperature and various pressures, recorded upon pressure increase and after total pressure release (top spectrum). The low frequency region has been suitably enhanced in order to improve visibility. The asterisk marks a band due to the methanol-ethanol mixture.

**Figure 2.** Pressure dependence of the Raman modes in DWCNTs. In the low frequency region (left panel) only the stronger and well-resolved peaks are plotted. The open (solid) symbols denote data taken for increasing (decreasing) pressure while solid lines are least square fittings.

**Figure 3.** (a) Integrated intensities of the $R_1$, $R_2$ (circles, outer tubes) and $R_4$ (squares, large inner tubes) radial bands normalized to the higher frequency RBM band ($R_5$ peak region, small inner tubes) as a function of pressure. (b) Pressure dependence of the full width at half maximum (FWHM) of the two strongest tangential modes $G_5$ and $G_6$, corresponding to inner and outer nanotubes, respectively. The open (solid) symbols denote data taken for increasing (decreasing) pressure in both panels while solid lines are drawn to guide the eye.

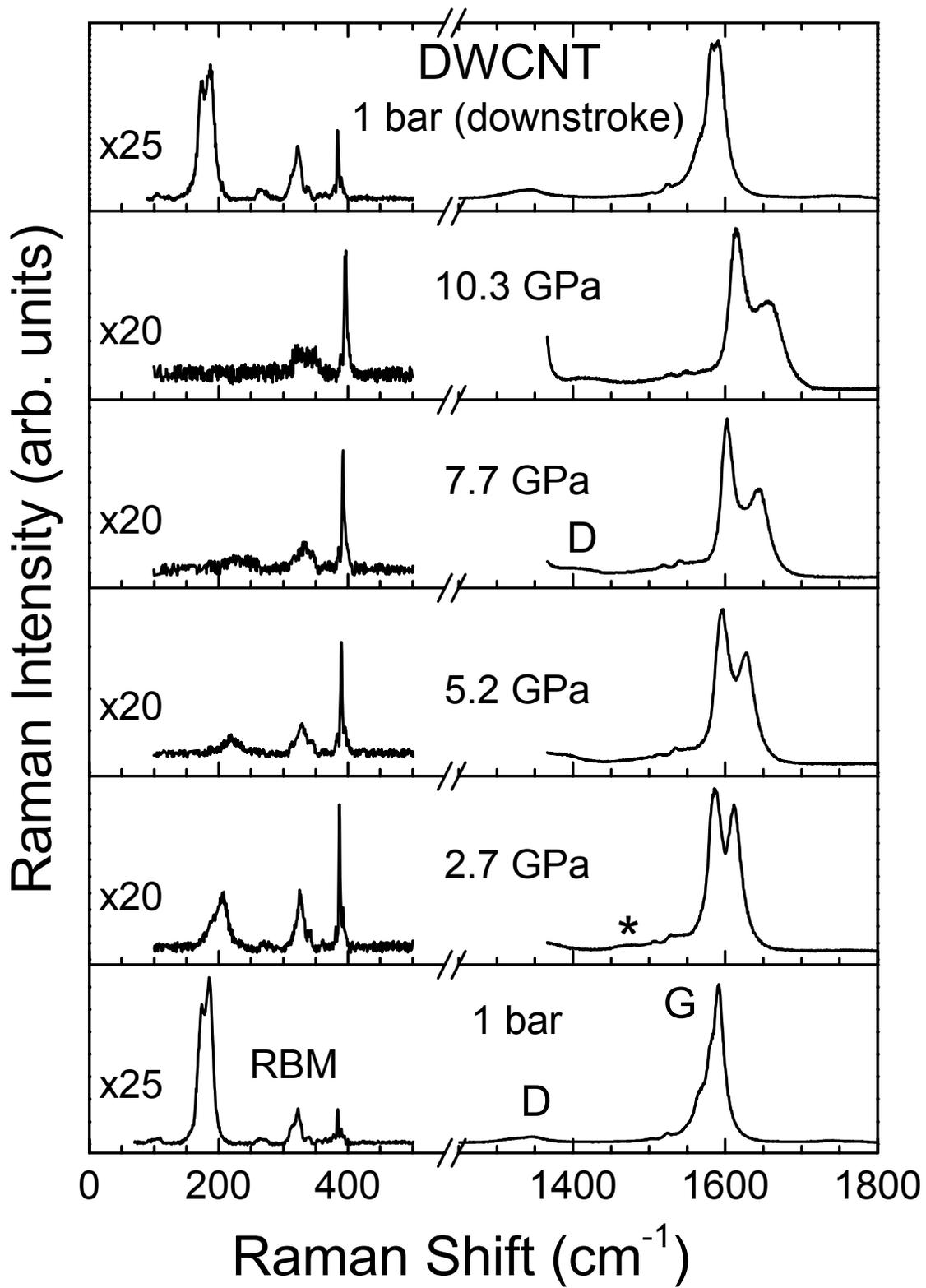

**Fig. 1**

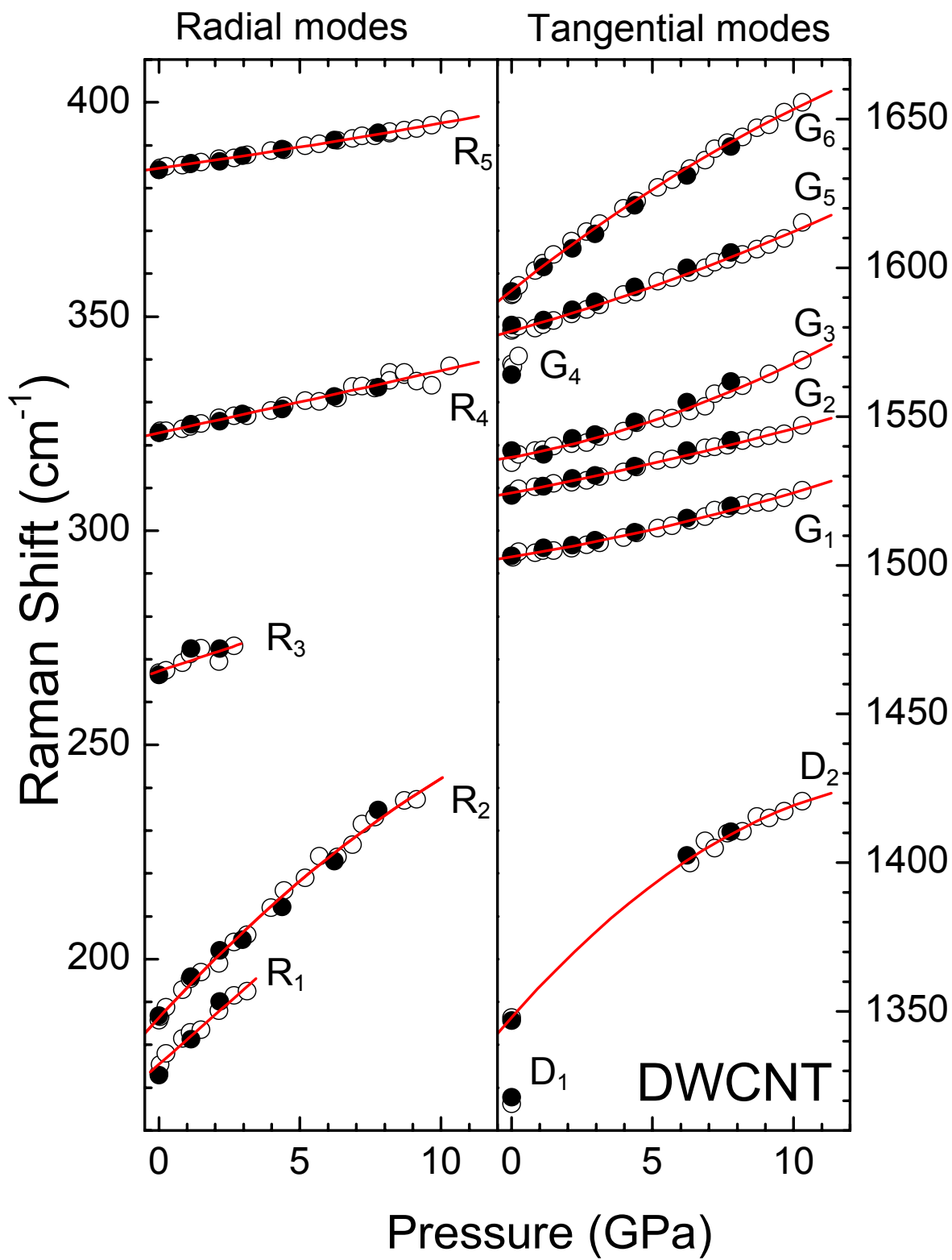

**Fig. 2**

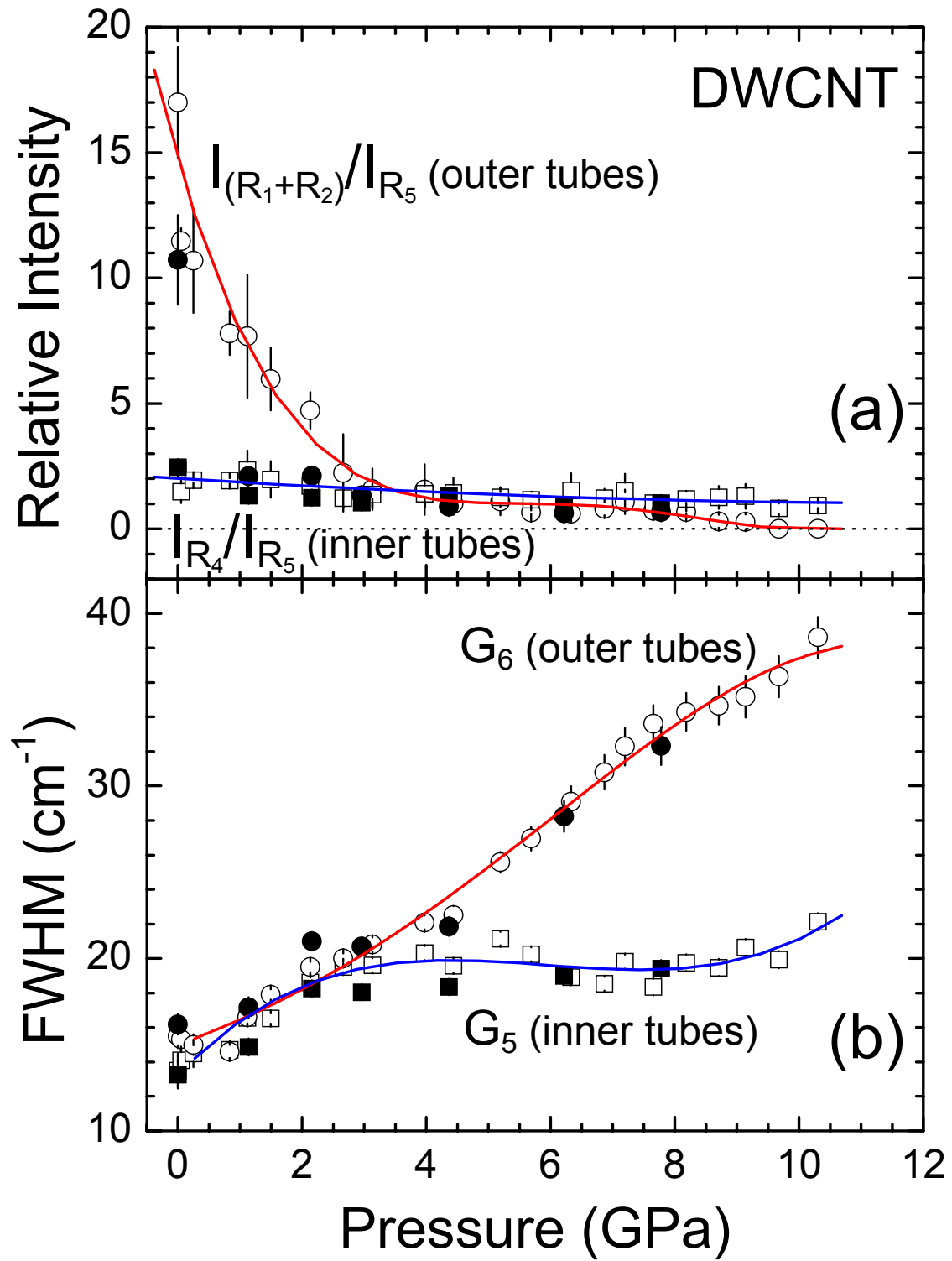

**Fig. 3**

**Table I.** The phonon frequencies and their pressure coefficients for the well resolved Raman peaks in DWCNTs.

| Mode | Parabolic fitting | | | Linear fitting |
|---|---|---|---|---|
| | $\omega_i$ (cm$^{-1}$) | $\partial\omega_i/\partial P$ (cm$^{-1}$/GPa) | $\partial^2\omega_i/\partial P^2$ (cm$^{-1}$/GPa$^2$) | $\partial\omega_i/\partial P$ (cm$^{-1}$/GPa) |
| $R_1$ | 175 | - | - | 5.8 |
| $R_2$ | 186 | 7.17 | -0.16 | 5.8 |
| $R_3$ | 267 | - | - | 2.2 |
| $R_4$ | 323 | - | - | 1.5 |
| $R_5$ | 384 | - | - | 1.1 |
| $D_1$ | 1320 | - | - | - |
| $D_2$ | 1348 | 10.71 | -0.36 | 7.1 |
| $G_1$ | 1503 | 1.51 | 0.06 | 2.1 |
| $G_2$ | 1524 | 1.80 | 0.04 | 2.2 |
| $G_3$ | 1536 | 1.75 | 0.14 | 3.1 |
| $G_4$ | 1567 | - | - | - |
| $G_5$ | 1579 | 2.65 | 0.07 | 3.3 |
| $G_6$ | 1592 | 7.50 | -0.14 | 6.1 |